\documentclass[twocolumn,aps,prl,showpacs,preprintnumbers,amsmath,amssymb,10pt]{revtex4-1}

\usepackage{graphicx}
\usepackage{dcolumn}
\usepackage{setspace}
\usepackage{bm}
\usepackage{multirow}

\begin{document}

\title{Investigating the dependence of collective dynamics on n/p asymmetry for light nuclei}

\author{R.~T. deSouza}
\email{desouza@indiana.edu}
\author{Varinderjit Singh}
\author{S. Hudan}
\affiliation{%
Department of Chemistry and Center for Exploration of Energy and Matter, Indiana University\\
2401 Milo B. Sampson Lane, Bloomington, Indiana 47408, USA}%

\author{Z. Lin}
\affiliation{%
Department of Physics and Center for Exploration of Energy and Matter, Indiana University\\
2401 Milo B. Sampson Lane, Bloomington, Indiana 47408 USA}%
{\affiliation{%
Department of Physics, Arizona State University, \\ 450 E. Tyler Mall, Tempe, AZ 85287-1504 USA}%

\author{C.~J. Horowitz}
\affiliation{%
Department of Physics and Center for Exploration of Energy and Matter, Indiana University\\
2401 Milo B. Sampson Lane, Bloomington, Indiana 47408, USA}%

\date{\today}

\begin{abstract}
  The dynamics present in the fusion of neutron-rich nuclei is explored through the comparison of experimental cross-sections at above-barrier energies
  with measurements of the interaction cross-section at
  relativistic energies. The increase of fusion dynamics with increasing neutron excess is clearly demonstrated. Experimental cross-sections are
  compared with the predictions of a Sao Paulo model using relativistic mean field density distributions and the impact of different interactions is explored.

\end{abstract}


\maketitle

Nuclei are extremely interesting quantal systems, which despite a limited number of constituent particles, manifest collective dynamics.
This collective dynamics is observed in many forms including the giant dipole resonance \cite{Bertrand76, Kobayashi89}, shape coexistence \cite{Heyde11}, and
fission \cite{Meitner39,Van_Huiz73}. Although typically associated with the structure and reactions of mid-mass and heavy nuclei, collectivity for very
light nuclei has recently been reported \cite{Morse18}. Nuclear fission and nuclear fusion provide examples in which collective
degrees of freedom undergo substantial change as the reaction proceeds. Of particular interest is the role of collectivity for
neutron-rich nuclei as for these nuclei the dependence of the dynamics on the asymmetry between the neutron and proton densities
can be probed.
Fusion reactions provide
a powerful means to assess the response of neutron-rich
nuclei to perturbation. As fusion involves the interplay of
the repulsive Coulomb and attractive nuclear potentials,
by examining fusion for an isotopic chain one probes the
neutron density distribution and how that density distribution evolves as the two nuclei approach and overlap \cite{Singh17,Vadas18}.
In the following manuscript we propose a
novel perspective for investigating the role
of collective dynamics in fusion. Moreover, using this new perspective we elucidate the dependence of the fusion
dynamics on n/p asymmetry for the first
time including the indication that for light nuclei fusion dynamics is enhanced with increasing neutron number.

Measurement of the interaction cross-section, $\sigma_{Int.}$ in high energy collisions is an effective means to invesitgate the
spatial extent of the matter distribution \cite{Ozawa96}.
The interaction cross-section in these measurements is simply defined as the total nuclear reaction cross-section resulting in a change of either the
atomic number (Z) or mass number (A) of the projectile. 
Systematic comparison of these cross-sections for lithium isotopes revealed the halo nature of $^{11}$Li \cite{Tanihata85a,Tanihata85b}. 
Presented in Fig.~\ref{fig:fig1} are the interaction cross-sections of carbon isotopes with a carbon target. Measurements for A$\geq$12
were made at E/A $\sim$ 900 MeV at GSI-Darmstadt and utilized the high-resolution fragment separator FRS to resolve the reaction products \cite{Ozawa01a, Ozawa01b}. These data, represented as solid symbols in Fig.~\ref{fig:fig1}, are supplemented by the results of
earlier measurements at the LBL Bevalac indicated by open symbols \cite{Ozawa96}. The overall trend observed is an approximately linear
increase in $\sigma_{Int.}$ with neutron excess, (N-Z). At the high incident energy that these experiments were conducted at, one
expects the sudden approximation to be valid. Hence, the measured interaction cross-section, $\sigma_{Int.}$ provides a direct measure of the extent of the matter distribution. Through comparison with a Glauber model, the rms matter radii of these nuclides has been
extracted \cite{Kanungo16}.

Closer examination of Fig.~\ref{fig:fig1} provides an indication of the impact of shell structure on 
$\sigma_{Int.}$.
The dependence of $\sigma_{Int.}$ on neutron excess for 12$\le$A$\le$14 is weak as is the dependence for
16$\le$A$\le$18. Between $^{14}$C and $^{16}$C one observes a jump in $\sigma_{Int.}$ from a value of $\sim$850 mb to
$\sim$1050 mb. This increase reflects the completion of the 1p$_{\frac{1}{2}}$ with N=8 and the population of the 1d$_{\frac{5}{2}}$ shell.
This observation is significant as it indicates that the shell structure of the neutron-rich isotopes is accessible through
measurement of $\sigma_{Int.}$ for an isotopic chain.

\begin{figure}
\includegraphics[scale=0.42]{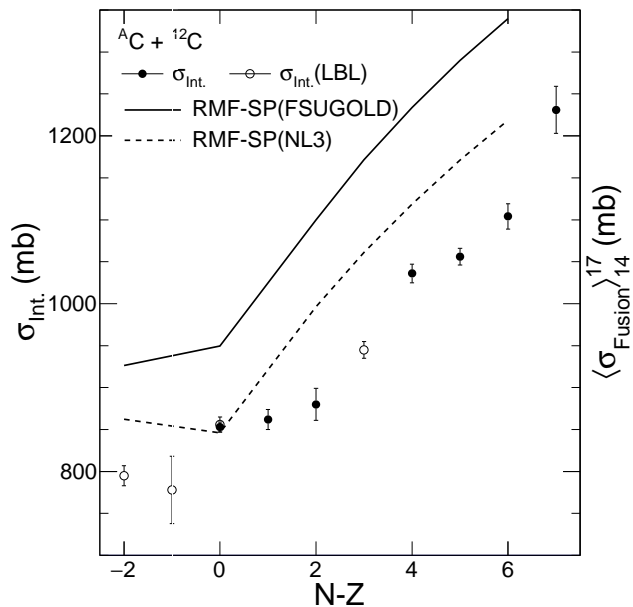}
\caption{\label{fig:fig1} Comparison of the interaction cross-section $\sigma_{Int.}$ for various carbon
  isotopes with the prediction of the average fusion cross-section at above-barrier energies using a RMF-SP model.
  See text for details.
}
\end{figure}

\begin{figure}
\includegraphics[scale=0.45]{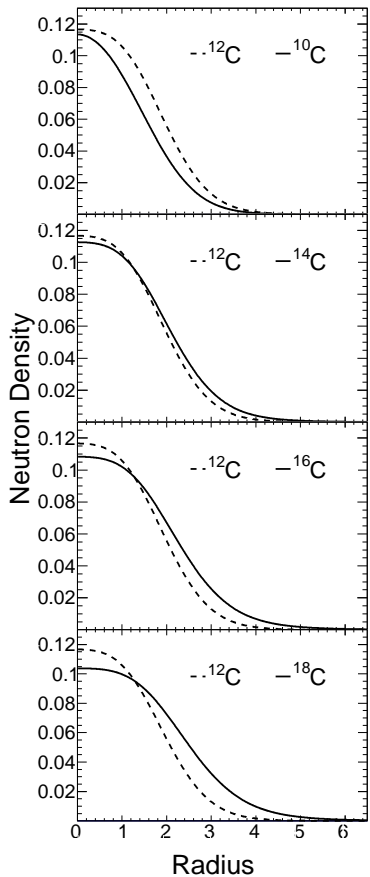}
\caption{\label{fig:fig2} Neutron density distributions predicted by the RMF model with the
 FSUGOLD interaction for carbon isotopes.
}
\end{figure}

Juxtaposed with the measured interaction cross-sections in Fig.~\ref{fig:fig1} are the results of calculations of the fusion cross-section at energies just above the fusion barrier. These calculations were performed at energies of E$_{CM}$ = 1-2 MeV/A and utilize the
Sao Paulo model for calculating the fusion cross-section \cite{Gasques04}. In this model the density distributions of the colliding nuclei are assumed to be frozen during the fusion process thus the calculated cross-sections reflect the size of the colliding nuclei. Moreover, use of a common target nucleus allows one to assess the change in the size of the projectile nucleus with increasing neutron
excess. The density distributions used in the Sao Paulo
calculations were determined using a relativistic mean
field (RMF) model \cite{Serot86,Ring96}. In order to investigate the sensitivity of the calculated cross-sections to the interaction used in the RMF calculations, the RMF calculations were performed using two sets of interactions FSUGOLD and
NL3 \cite{Shen11}. In contrast to the widely used NL3 interaction,
the FSUGOLD corresponds to a softer interaction \cite{Fattoyev10}. To facilitate the comparison of the fusion cross-sections predicted by the RMF-SP
model with $\sigma_{Int.}$ we have calculated the average fusion cross-section over the interval 14 MeV$\le$E$_{CM}$$\le$17 MeV and
designate this quantity $<$$\sigma_{fusion}$$>$$_{14}^{17}$. Although the calculations with the FSUGOLD interaction manifest a
consistently larger fusion cross-section, nonetheless both RMF-SP calculations exhibit a similar
dependence on neutron number. For N$<$Z the value of $<$$\sigma_{fusion}$$>$$_{14}^{17}$ is approximately constant
while for N$>$Z it increases approximately linearly with (N-Z). The slope of the predicted cross-sections for the two interactions shown
is to first order
the same indicating that while the absolute size of the nucleus depends on the interaction used in the RMF model, the
increase in size with increasing N is relatively insensitive to the interaction utilized.
Moreover for N$>$Z,
the slope of the predicted above-barrier fusion cross-section 
is very similar to that for $\sigma_{Int.}$. This similarity of the two slopes arises
from the fact that the $\sigma_{Int.}$ measures the size of the nucleus and the RMF-SP with the frozen density distributions is
intrinsically related to the same quantity. {\em As such, the quantity $\sigma_{Int.}$ provides a key reference
from which to examine fusion dynamics.}

\begin{figure}
\includegraphics[scale=0.45]{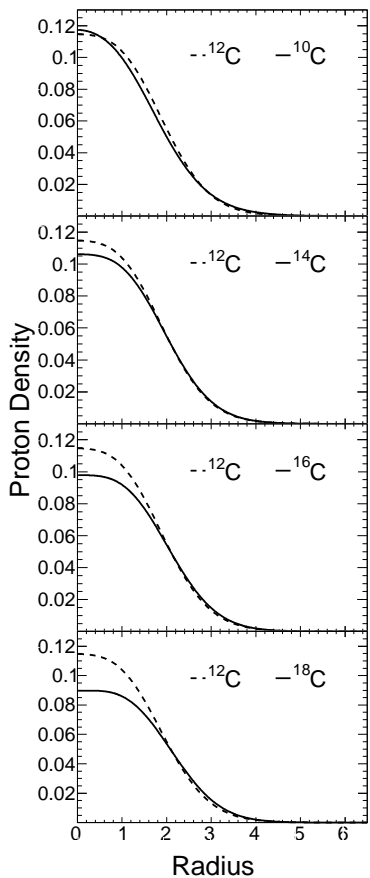}
\caption{\label{fig:fig3} Proton density distributions predicted by the RMF model with the
 FSUGOLD interaction for carbon isotopes.
}
\end{figure}

In Fig.~\ref{fig:fig2} the density distributions for neutrons predicted by the
RMF model for the various carbon isotopes are displayed. The density distributions for N$\ne$Z are compared with that of $^{12}$C for reference. As expected,
the tail of the neutron density distribution extends further for the
more neutron-rich the isotope. The value of calculating these neutron density distributions for an isotopic chain
lies in the ability to examine the systematic dependence on neutron number.
The evolution of the nuclear size on neutron number may have different sensitivity to the model uncertainties as compared to the absolute size.
Presented in
Fig.~\ref{fig:fig3} are the density distributions for protons predicted by the RMF model for the different isotopes of carbon. While
the distributions are all quite close as might be expected, as the isotope becomes more neutron-rich the tail of the proton
distribution extends slightly further out. This change in the proton distribution is due to the attractive nuclear force of the
valence neutrons. The dependence of the charge radii on neutron number for the carbon isotopic chain has recently been
determined through
measurement of the charge changing cross-section \cite{Kanungo16}.

\begin{figure}
\includegraphics[scale=0.45]{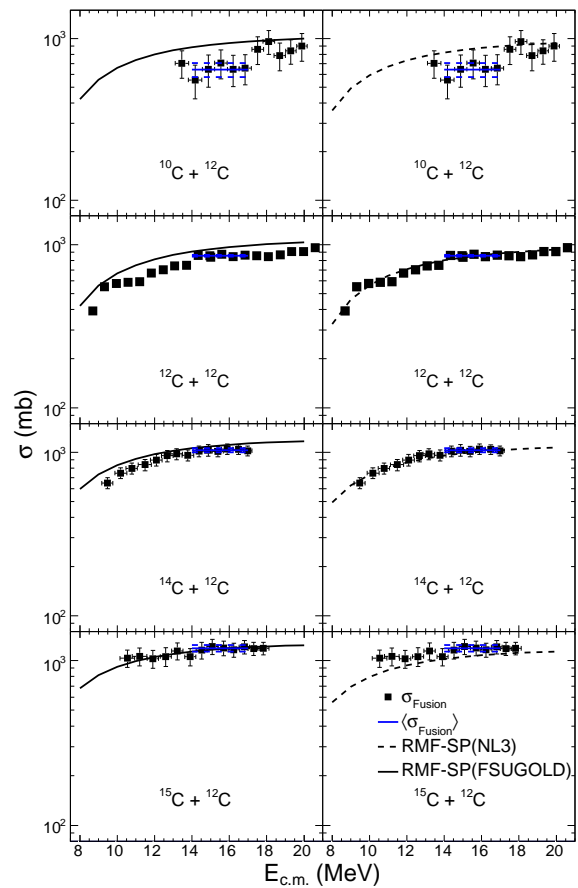}
\caption{\label{fig:fig4} Fusion excitation functions for $^{10-15}$C + $^{12}$c. Experimental data are compared with the results
  of a RMF-SP model using FSUGOLD (left column) and NL3 (right column). The fusion excitation function for $^{13}$C (not shown) is comparable to that of $^{14}$C \cite{Calderon15, Carnelli_thesis14}.
}
\end{figure}

To investigate the evolution of fusion dynamics with increasing neutron number we examine the fusion cross-section for $^A$C+ $^{12}$C at near barrier energies. Using a novel active target approach the fusion excitation functions for these reactions was
measured by the ANL group \cite{Carnelli14}. This active target approach is particularly well suited to studying reactions
with low intensity beams and allowed measurement of the fusion excitation function with
beam intensities as low as 500 ions/s. Depicted in Fig.~\ref{fig:fig4} (both columns) are the fusion  
cross-section data for $^{10-15}$C.
The data for $^{10,15}$C have been taken from \cite{Carnelli15}. Using the same approach though at higher beam
intensity, 
$^{12,14}$C cross-sections were also measured \cite{Calderon15,Carnelli_thesis14} and are shown in Fig.~\ref{fig:fig4}.
The fusion excitation
functions observed for  $^{12,13}$C + $^{12}$C are in good agreement with those published in the literature \cite{Kovar79,Dayras76}.
The measured excitation
functions manifest the expected dependence indicative of a barrier driven process. The experimental data are compared with the results
of the RMF-SP model with the FSUGOLD and NL3 interactions depicted in the left and right columns respectively. While in the case
of $^{10}$C the models overpredict the experimental results, in the remainder of the cases the agreement is reasonable. Close
comparison of the left and right columns indicates that the calculations with FSUGOLD consistently predict larger cross-sections
than those with NL3, consistent with the observation in Fig.~\ref{fig:fig1}. It is noteworthy though that this increase in the cross-section for FSUGOLD as compared to NL3 is typical of the entire above-barrier regime. In Fig.~\ref{fig:fig4} the blue bar indicates the
value of the average cross-section as well as the energy interval over which the average was calculated. For $^{12-15}$C the average cross-section is clearly representative of the above-barrier cross-section. In the case
of $^{10}$C, however there is significant variation in the measured cross-sections and the average
cross-section calculated is more sensitive to the choice of energy interval.

In Fig.~\ref{fig:fig5} the average above-barrier fusion cross-sections for the carbon isotopic chain are compared with
$\sigma_{Int.}$. One observes that for $^{12}$C the fusion cross-section and $\sigma_{Int.}$ are essentially the same. For N$>$Z
however, the fusion cross-section depends more strongly on neutron excess than $\sigma_{Int.}$ does. Since the dependence of
$\sigma_{Int.}$
on increasing neutron number indicates the inherent growth in the size of the neutron density distribution with
increasing neutron number, the cross-section in the case of fusion, above that of $\sigma_{Int.}$
reflects the impact of dynamics in the fusion process. The quantity ($<$$\sigma_{fusion}$$>$$_{14}^{17}$-$\sigma_{Int.}$)
can be viewed as a measure of the fusion dynamics.
Moreover, the increase in this quantity, dictated by the larger slope for fusion as compared to $\sigma_{Int.}$,
indicates that this dynamics evolves with increasing neutron excess.

It is also instructive to compare the behavior of the fusion data with the results of the RMF-SP model. The experimental data
indicates a stronger dependence on neutron excess than the RMF-SP model independent of the interaction chosen. This comparison  thus
also indicates an increased role for dynamics with increasing neutron excess. Thus, the increased role of fusion dynamics with neutron excess is realized in two independent ways. Juxtaposition of the $<$$\sigma_{fusion}$$>$$_{14}^{17}$
with $\sigma_{Int.}$, namely a data to data comparison, indicates enhanced
fusion dynamics with increasing neutron excess.
This result is supported by the comparison of $<$$\sigma_{fusion}$$>$$_{14}^{17}$ with the cross-sections predicted by the RMF-SP model.

Although the span of neutron excess for the fusion
data presented is presently limited, the new generation of radioactive beam facilities allows one to 
extend these measurements to even more neutron-rich carbon isotopes.
Measurement of fusion with beams of $^{16,17}$C and possibly $^{18,19}$C at FRIB \cite{FRIB} is envisoned. Similar measurement for the
oxygen isotopic chain extending nearly to the neutron-drip line is also possible.

Examination of the fusion cross-section at above-barrier energies for an isotopic chain is a powerful tool.
Comparison of fusion cross-sections just above the barrier with
the interaction cross-section, $\sigma_{Int.}$, at high energies where the sudden approximation is valid allows extraction of not just
the fusion dynamics but the dependence of the dynamics on neutron excess. Investigating this dynamics for the most neutron-rich
nuclei accessible could provide valuable insight into the dynamics of extremely asymmetric nuclear matter.

\begin{figure}
\includegraphics[scale=0.42]{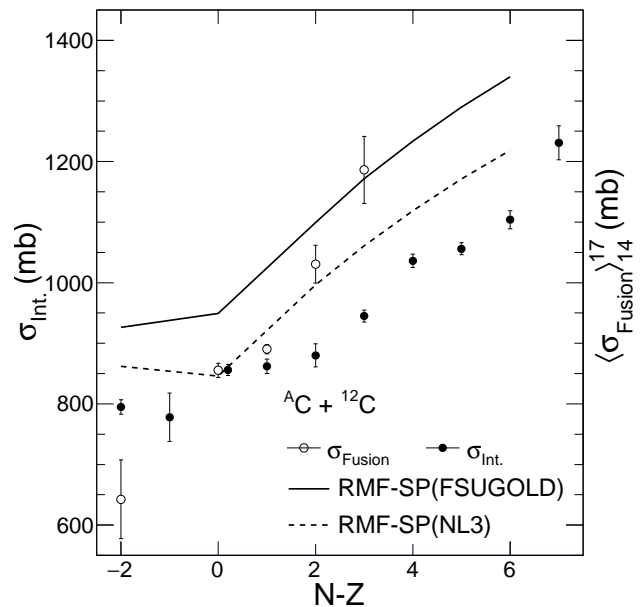}
\caption{\label{fig:fig5}
  Comparison of the average fusion cross-section $<$$\sigma_{fusion}$$>$$_{14}^{17}$ with the 
  interaction cross-section $\sigma_{Int.}$ for various carbon
  isotopes. Also shown are the predictions of the average fusion cross-section using a RMF-SP model.
}
\end{figure}

\begin{acknowledgments}
This work was supported by the U.S. Department of Energy under Grant No. 
DE-FG02-88ER-40404 (Indiana University).
CJH is supported in part by U.S. DOE grants DE-FG02-87ER40365 and DE-SC0018083.
ZL gratefully acknowledges support from National Science Foundation under PHY-1613708 (Arizona State University).
\end{acknowledgments}

%


\end{document}